\documentclass[10pt,a4paper]{article}
\usepackage{graphpap,epsfig,amssymb,graphicx,textcomp}
\usepackage[utf8]{inputenc}
\usepackage[english]{babel}
\begin{document}
\begin{center}{\Large{\bf 
	Role of anisotropy to the compensation 
	in the Blume-Capel trilayered
	ferrimagnet}}
\end{center}

\vskip 1cm

\begin{center}{\it Muktish Acharyya}\\
{\it Department of Physics, Presidency University}\\
{\it 86/1 College street, Calcutta-700073, India}\\
{E-mail:muktish.physics@presiuniv.ac.in}\end{center}

\vskip 2cm

\noindent {\bf Abstract:}
The trilayered Blume-Capel ($S=1$) magnet with 
nearest neighbour intralayer ferromagnetic
and nearest neighbour 
interlayer antiferromagnetic interaction is studied by Monte
Carlo simulation. Depending on the relative interaction strength
and the value of anisotropy the critical temperature (where all the
sublattice magnetisations and consequently 
the total magnetisation vanishes) and
the compensation temperature (where the total magnetisation vanishes
for a special combination of nonzero sublattice magnetisations) 
are estimated. The comprehensive phase diagrams 
with lines of critical temperatures and compensation temperatures
for different parameter 
values are drawn. 

\vskip 1cm


\newpage

\noindent {\bf I. Introduction:}

To study the magnetocaloric effects\cite{phan}, magneto-optical 
recording
\cite{connell} and the giant magnetoresistance\cite{camley}, 
the ferrimagnetic
materials are widely used for experimental  and theoretical studies.
The thermomagnetic recording \cite{phan}
device requires the strong temperature
dependence of the coercive field. Some ferrimagnetic materials shows
compensation (the total magnetisation vanishes for nonzero sublattice
magnetisations) at room temperatures where the coercive field is
strongly dependent on the temperature. The trilayered ferrimagnetic materials
show an interesting phenomenon, called compensation. Below the critical
temperature
(where each of all sublattice magnetisations as well as the total 
magnetisation vanishes),
 for particular combinations of ferromagnetic and 
antiferromagnetic interaction strengths, the total magnetisation vanishes
even for nonzero value of each of sublattice magnetisations. It has been reported
that near the compensation, the system shows diverging coercivity, and
a good choice for thermomagnetic and magneto-optic 
recording\cite{sheih}. Due to this modern technological importance the
study of compensation phenomena became quite interesting to the 
experimental and theoretical researchers. With the practical realisation
of the layered magnetic materials, such as 
bilayer\cite{stier}, trilayer\cite{smits,leiner}
and multilayer\cite{kepa,chern,sankowski,chung}, 
theoretical inversigations are required for
better understanding of compensation phenomena.

Since, the exact theoretical treatments 
are not adequately available in the literature, 
the approximation methods are applied to study such complex
compensation phenomena in the magnetic model systems. The trilayer spin-${{1} \over {2}}$
ferrimagnets are the prototypes to study\cite{diaz} such effects.
The Monte Carlo approach was employed\cite{naji} to study the 
compensation in
Ising trilayered magnetic models.

The magnetic anisotropy plays important role to change the critical and compensation 
temperatures in the magnetic materials.
The dependences of critical and compensation temperatures
on the crystal field anisotropy was observed \cite{cardona}
in mixed spin $({{5} \over {2}}, {{3} \over {2}})$ Ising antiferromagnetic core-shell nanowire. The critical and compensation temperatures were found \cite{cardona2} 
to depend significantly
on the crystal field anisotropy in mixed spin $({{5} \over {2}}, {{3} \over {2}})$
Ising ferrimagnetic graphene layer. The single site anisotropy plays crucial role
in the thermodynamic behaviours of magnetic spin systems. The spin-1 anisotropic
(easy-axis single ion type) Heisenberg ferromagnet is studied\cite{yangwang} 
by Green function diagramatic technique which shows the temperature expansion of
magnetisation, Dyson's $T^4$ correction to the first Born approximation, along with
a series term led by $T^2 e^{-{\beta} D}$ for the single ion anisotropy $D$.

The double compensastion temperatures are found
\cite{fadil2} in a mixed spin
(${{7}\over{2}},1$) antiferromagnetic
ovalene nanostructured system studied by MC
simulation. The
Monte Carlo methods were employed to study the Blume-Capel 
bilayered graphene structure with RKKY interactions. 
It was observed\cite{fadil3} 
that the transition temperature increases with 
decreasing the number of
nonmagnetic layers. However, the behaviours of Blume-Capel trilayer
is not yet studied by MC method.

Although the Blume-Capel (BC) model\cite{blume,capel,emery} was originally
introduced to analyse the thermodynamic behaviours of $\lambda$-transition
in the mixture of He$^3$-He$^4$, it is widely used to study the
bicritical/tricritical behaviours in various phase transitions. The 
nature (discontinuous/continuous) of the phase transition and the existence
of tricritical point in face centered cubic BC model 
was studied by
high temperature series extrapolation techniques\cite{stauffer} and Monte Carlo
simulation\cite{jain}.
The tricritical behaviour
\cite{deserno,lara} in the BC model was studied by Monte carlo simulation.

The meanfield approximation was employed to study\cite{general} the
general spin BC model.
The meanfield solution was obtained\cite{santos} 
in BC model (infinite range ferromagnetic interaction) with random crystal
field also. The method of effective field theory 
was used\cite{yuksel} to study the
effects of random crystal field in the BC model.
The wetting transition in BC model was 
studied\cite{albano} by MC simulation.

The BC model exhibits the competing metastability. 
It should be mentioned here
that dynamic Monte Carlo and numerical transfer matrix method \cite{fiig}
were employed to study the competing metastability in the BC model.
The behaviours of the competing metasatble 
states at infinite volume are studied  
\cite{manzo} in dynamic BC model. The metastable and unstable 
states are obtained
\cite{ekiz} by cluster variation and path probability method. However,
no study was found to consider the compensation in the BC trilayerd
magnetic model systems. 

{\it What kind of behavious are expected in the Blume-Capel trilayerd 
ferrimagnet ? How does the anisotropy affect the compensation 
temperature ?} 
To address these questions,
in this aricle, the equilibrium behaviours of 
critical and compensation behaviours and the dependence of these two
temperatures on the single site magnetic anisotropy ($D$), are studied by Monte Carlo simulation.
This paper is organised as follows: the Blume-Capel ($S=1$) model, with a brief description of applied Monte Carlo technique, is described in the section
II, the numerical results are reported in section III and the paper ends with summary
in section IV.

\vskip 0.5cm

\noindent {\bf II. Model and simulation:}

The energy of such a Blume-Capel ($S=1$) trilayer is represented
by the following Hamiltonian,

\begin{equation}
	H = -{J_{aa}}\sum_{<ij>} S_i^zS_j^z 
	 -{J_{bb}}\sum_{<ij>} S_i^zS_j^z 
	 -{J_{ab}}\sum_{<ij>} S_i^zS_j^z 
	+D\sum_i (S_i^z)^2 
\end{equation}

\noindent where, $S_i^z$ represents the z-component of the Spin ($S=1$) at any position (i-th lattice site).
The values of $S_i^z$  
may be any one of -1,0 and +1. The first term represents the 
contribution to the energy due to 
nearest neighbour ferromagnetic ($J_{aa} >0$) 
interactions among the spins in top (A) layer and
the same in the bottom (A) layer. 
The second term represents the contribution to the energy due to the
nearest neighbour interactions among the spins in the middle (B) layer.
$J_{bb} > 0$ is the ferromagnetic nearest neighbour interaction
between the spins in the middle (B) layer. 
The contribution to the energy due to nearest neighbour 
inter-layer (A-B) antiferromagnetic ($J_{ab} <0$)
interaction is represented by the third term.
The summations in all these three terms are considered only over distinct
pairs to avoid any overcounting.
Finally, the fourth term is the contribution to the energy due to
the single site magnetic anisotropy, where $D$ is the strength of anisotropy. The periodic
boundary conditions are applied in both directions of each layer and such a trilayered system is kept in open boundary condition. This completes the description of the model.

In the simulation, a trilayered system of $L=100$ is considered. The
equilibrium configuration at any particular temperature 
($T$) was achieved just
by cooling the system slowly (with small change in temperature
$\Delta T$) from a high temperature disordered
state of random spin configurations. The high temperature random 
initial spin
configuration was generated in such a way that the system contains almost
equal numbers of $S_i^z = +1, 0$ and -1, distributed randomly. 
In such a configuration, all the sublattice 
magnetisations (for each of the three layers)  
and consequently the total magnetisation of the whole system vanishes. 
Now a high value of the temperature is considered. The spin ($S^z_i$) at any site
(i-th site) of the
system has been updated
randomly using Monte Carlo method with
the Metropolis formula\cite{springer}
\begin{equation}
	P(S_i^z(initial) \rightarrow S_i^z(final))
	={\rm Min}[1,{\rm exp}(-{{\Delta E} \over {kT}})]
\end{equation}
\noindent where $k$ is the Boltzmann constant.  In such way, $3L^2$ numbers of such
random updates of the spins are done and considered as the unit of time
(MCSS, Monte Carlo Step per Spin) in the simulation. In the present 
simulational study, $12 \times 10^5$ MCSS are considered, where the 
initial (transient) $6\times10^5$ MCCS were discarded and the quantities
are calculated by averaging over next $6\times10^5$ MCSS. Some results
are checked with smaller length of simulation and no significant changes
were observed. In that spirit, it was assumed that the system has reached
the equilibrium configuration of that given temperature ($T$). Now a lower
temperature (with $\Delta T =0.05$) is considered and the peresent spin 
configuration was used as the initial starting configuration for that 
lower temperature ($T-\Delta T$). In this way, the macroscopic quantities are calculated
for different temperatures. Assuming the ergodicity, the time average
serves the purpose of evaluating the ensemble average. 
The following quantities are calculated:
The sublattice magnetisations 
$m_{top}=<{{1}\over{L^2}}\sum_i S_i^z>; i ~~\forall ~~{\rm top (A) ~~layer}$,  
$m_{mid}=<{{1}\over{L^2}}\sum_j S_j^z>; j ~~\forall ~~{\rm middle (B) ~~layer}$  
and
$m_{bot}=<{{1}\over{L^2}}\sum_n S_n^z>; n ~~\forall ~~{\rm bottom (A) ~~layer}$  
the total magnetisation 
$m_{tot}^{'} = (m_{top}+m_{mid}+m_{bot})/3$, $m_{tot}=3m_{tot}^{'}$ and the susceptibility
$C=L^2{{J_{bb}} \over {kT}} <(m_{mid} -  
{{1}\over{L^2}}\sum_j S_j^z)^2>; j \forall {\rm middle (B) ~~layer}$, where
in all cases $<...>$ stands for the time average over $6\times10^5$ MCSS.
The temperature is measured in the unit of ${{J_{bb}} \over k}$.

\vskip 0.5cm

\noindent {\bf III. Results:}

The sublattice magnetisations of all three layers, the total 
magnetisation and the susceptibility are studied as functions 
of temperature. The values of $J_{bb}=1.0$ and $J_{ab}=-0.5$ 
are kept fixed throughout the study. Only the values of $J_{aa}$ and $D$
are varied. 

\begin{figure}[h]
\begin{center}
\begin{tabular}{c}
	(a)
\resizebox{7cm}{!}{\includegraphics[angle=0]{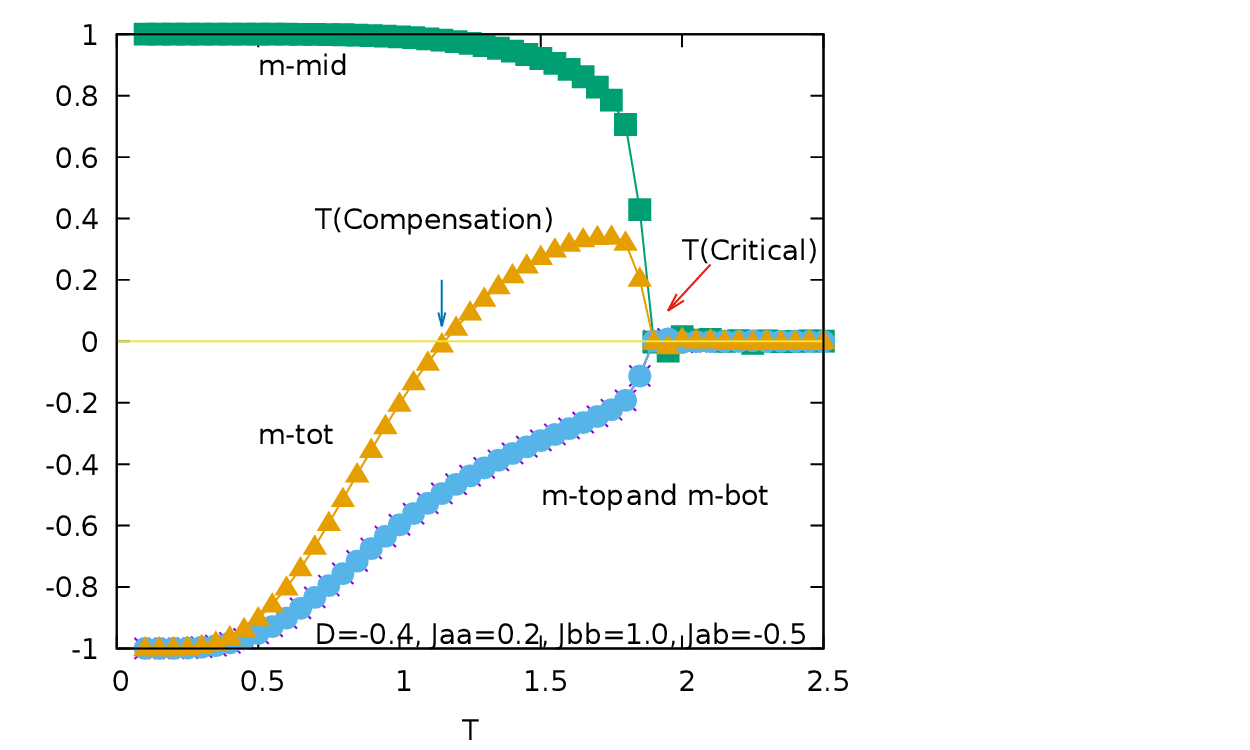}}
	(b)
\resizebox{7cm}{!}{\includegraphics[angle=0]{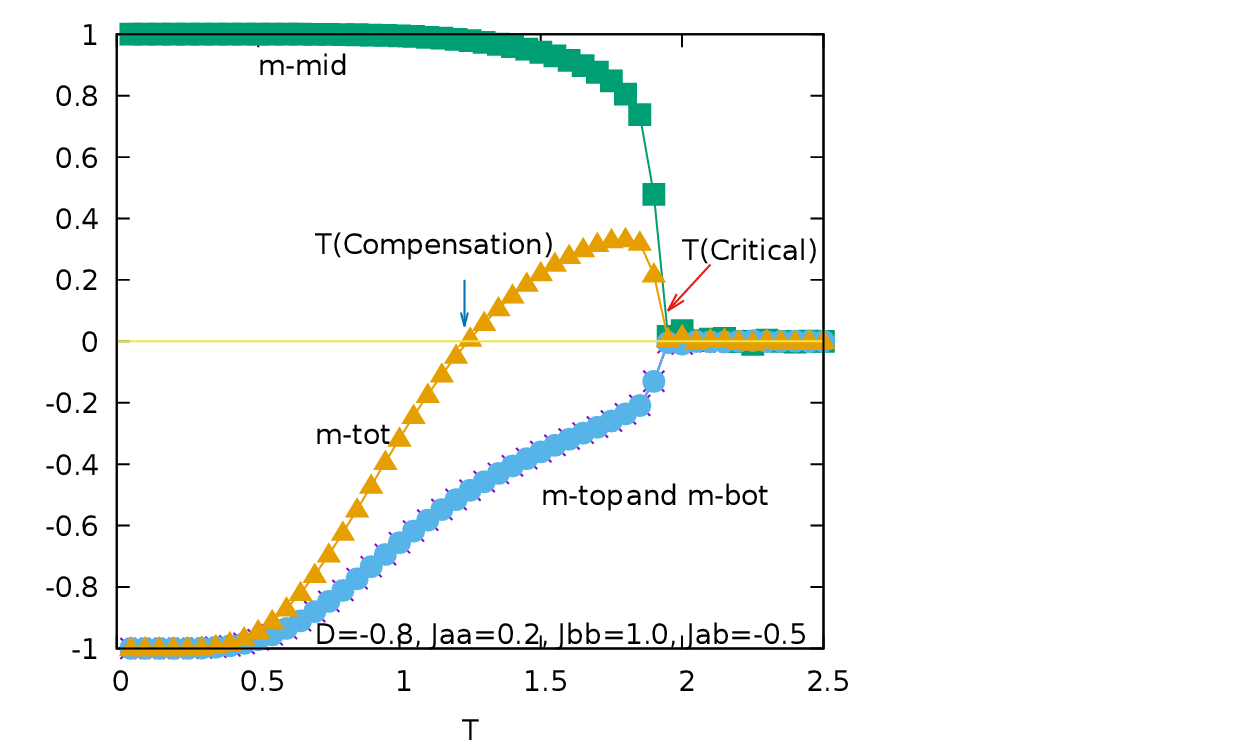}}
\\
	(c)
\resizebox{7cm}{!}{\includegraphics[angle=0]{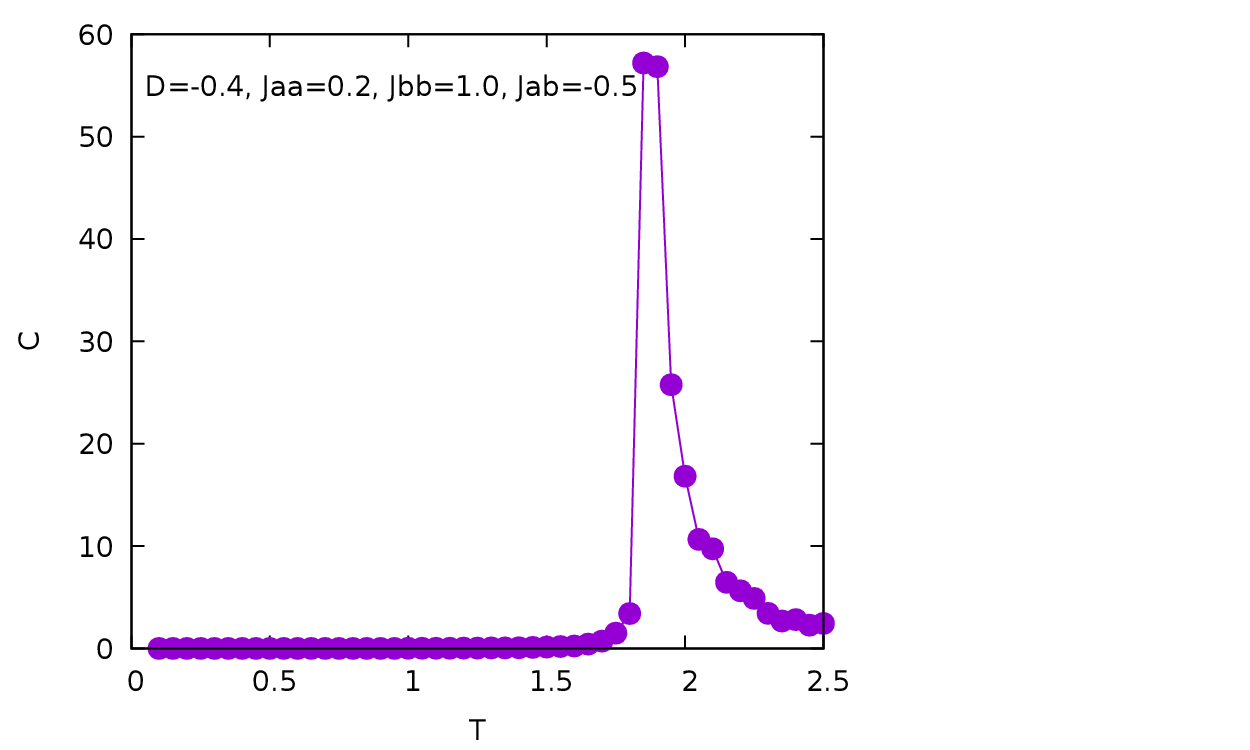}}
	(d)
\resizebox{7cm}{!}{\includegraphics[angle=0]{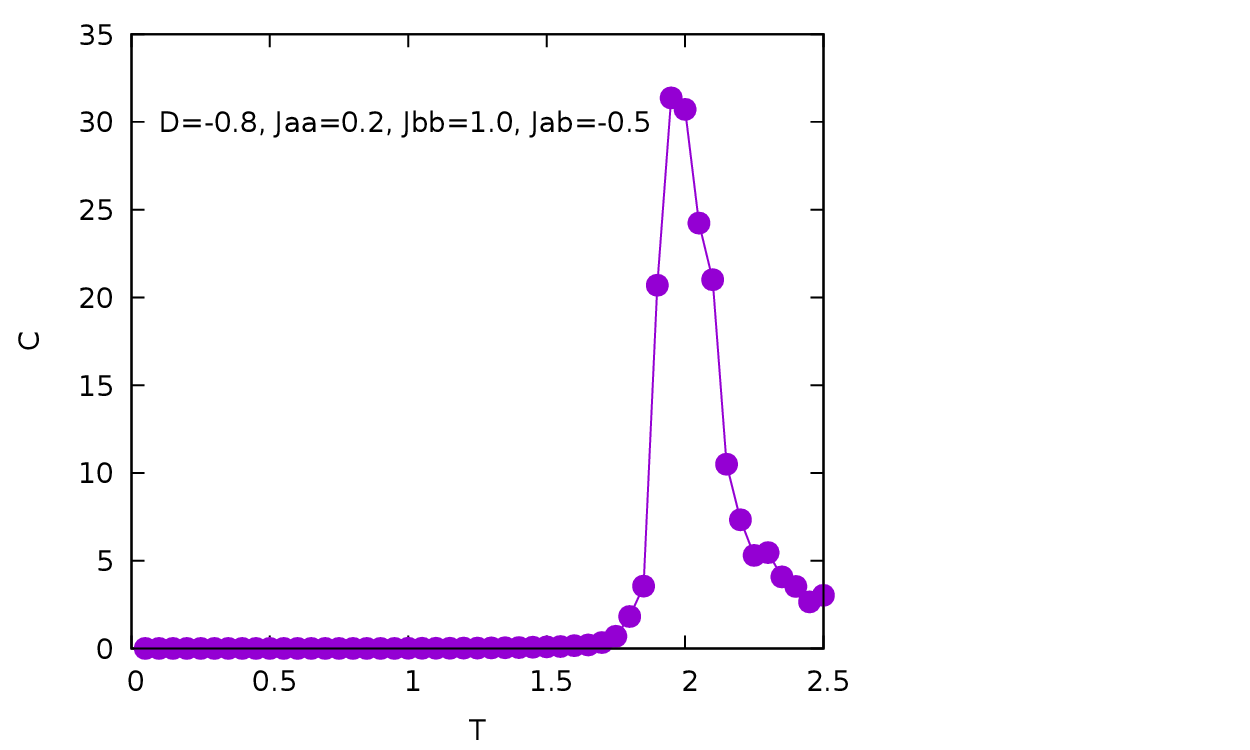}}
          \end{tabular}
\caption{The sublattice magnetisations of different layers and the total magnetisationare plotted against the temperature. The corresponding susceptibilities are
also plotted against the temperature of the system.
}

	\label{mT10.2}
\end{center}
\end{figure}

{\it What would happen for low $J_{aa}$, say $J_{aa}=0.2$ ?} 
In Fig-\ref{mT10.2}a the sublattice magnetisations, total 
magnetisation are shown for $J_{aa}=0.2$ and $D=-0.4$. The critical 
temperature $T_{critical}$ is also marked where all the sublattice 
magnetisations ($m_{top}=m_{mid}=m_{bot}=0$) and the total magnetisation
($m_{tot}=0$) vanish. Below the critical
temperature, there exists a temperature, so called compensation
temperature ($T_{compensation}$) 
where the total magnetisation vanishes ($m_{tot}=0$) for
nonzero values of sublattice magnetisations 
($m_{top} \neq 0, m_{mid} \neq =0, m_{bot} \neq 0$)
of all three layers.
For some other value of $D=-0.8$, the $T_{critical}$ and 
$T_{compensation}$ change, as shown in Fig-\ref{mT10.2}b. The critical
temperature $T_{critical}$ was measured from the position of the
maximum of the susceptibility $C$ plotted against temperature (shown
in Fig-\ref{mT10.2}a and Fig-\ref{mT10.2}b). However, the compensation
temperature was measured by linear interpolation in the region where
the total magnetisation changes sign below the critical temperature.
Since, the interval ($\Delta T$) of temperature for cooling the
system is equals to 0.05, the size of the maximum error in estimating
the critical and compensation temperature is 0.1. From Fig-\ref{mT10.2}
it is clear that both the compensation temperature $T_{compensation}$
and the critical temperature $T_{critical}$ decreases as the absolute value
of the anisotropy
$D$ decreases.

This compensation phenomenon can be realised as follows: if the intra
layer ferromagnetic interaction strength were chosen equal in all layers
and with a fixed inter layer antiferromagnetic interaction, the trilayered
system would exhibit almost equal magnitude of sublattice magnetisation.
However, top layer and bottom layer would show the same sign of 
sublattice magnetisation and middle layer would show 
different sign of sublattice magnetisation. As a result, the system would
show the total magnetisation (essentially the sublattice magnetisation
of any one of top and bottom layer) and only the critical temperature
would be found. Below the critical temperature no such temperature
was observed where the total magnetisation could vanish. But, if the
intra layer ferromagnetic interaction strength of top and bottom layers
is relatively weak in comparison to that for the middle layer, 
the sublattice magnetisation in top and bottom layers would be smaller
in magnitude (having same sign of course) than that of the middle
layer (having opposite sign). As a result, below the critical temperature,
a temperature could be found where the net magnetisation vanishes.
What will be the role of anisotropy $D$ ? Large negative value of $D$
(in equation-1), will map the Blume-Capel model in spin-1/2
Ising model. Relatively, weak negative $D$ will produce a few number of
$S^z=0$. For positive $D$ the number of $S^z=0$ will increase.
The sites having $S_i^z=0$ will not contribute to the sublattice magnetisations. So, by
changing the value of anisotropy $D$, one can control the 
value of the magnitude of sublattice magnetisation at any fixed 
temperature. 

\begin{figure}[h]
\begin{center}
\begin{tabular}{c}
\resizebox{7cm}{!}{\includegraphics[angle=0]{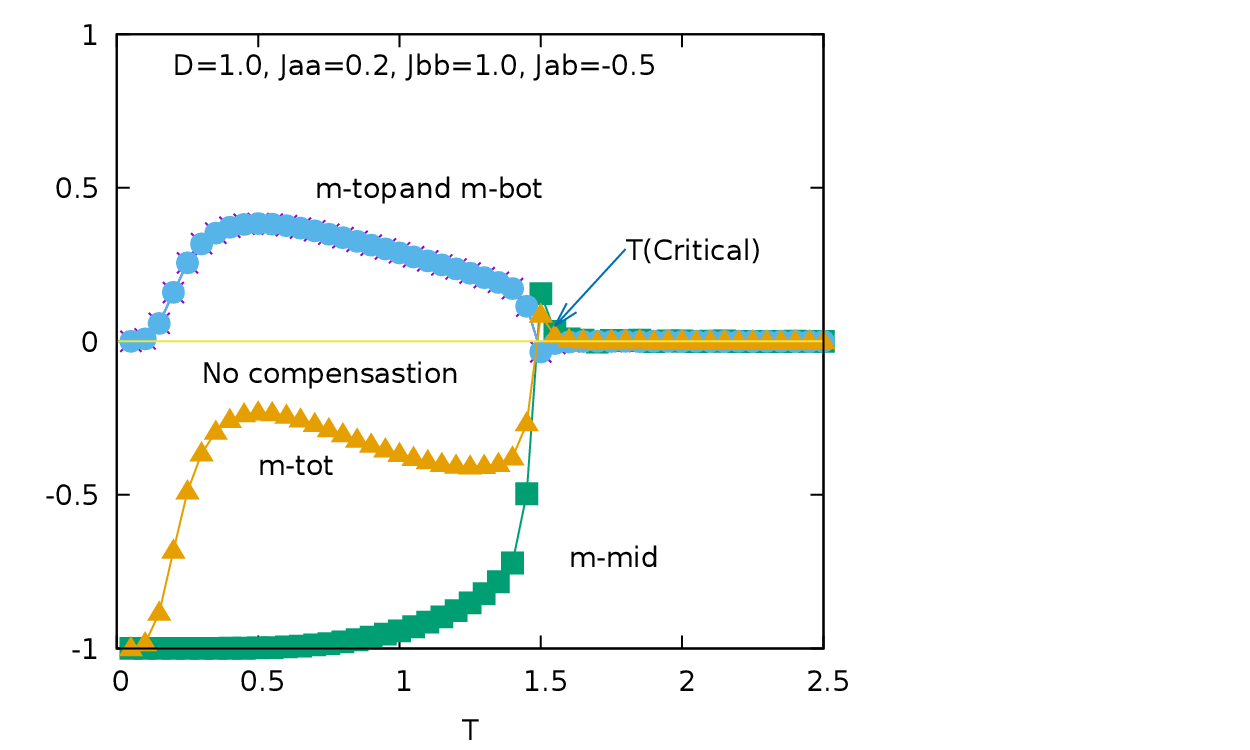}}
          \end{tabular}
\caption{The sublattice magnetisations of different layers and the 
	total magnetisationare plotted against the temperature. 
}

	\label{mT20.2}
\end{center}
\end{figure}

The compensation phenomenon was found to disappear for larger and positive
value of $D=1.0$ and $J_{aa}=0.2$. This is shown in Fig-\ref{mT20.2}. In this case, the
number of $S_i^z=0$ is such that it is incapable of yielding the
compensation.

\begin{figure}[h]
\begin{center}
\begin{tabular}{c}
	(a)
\resizebox{7cm}{!}{\includegraphics[angle=0]{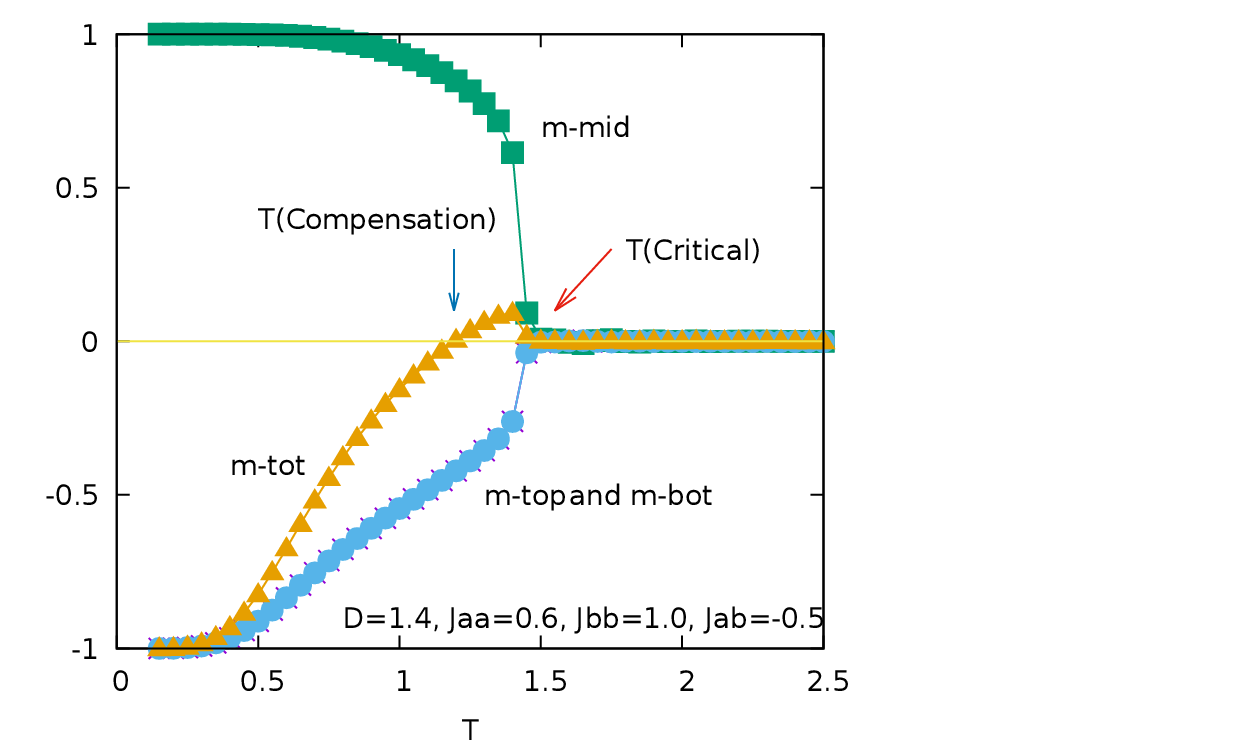}}
	(b)
\resizebox{7cm}{!}{\includegraphics[angle=0]{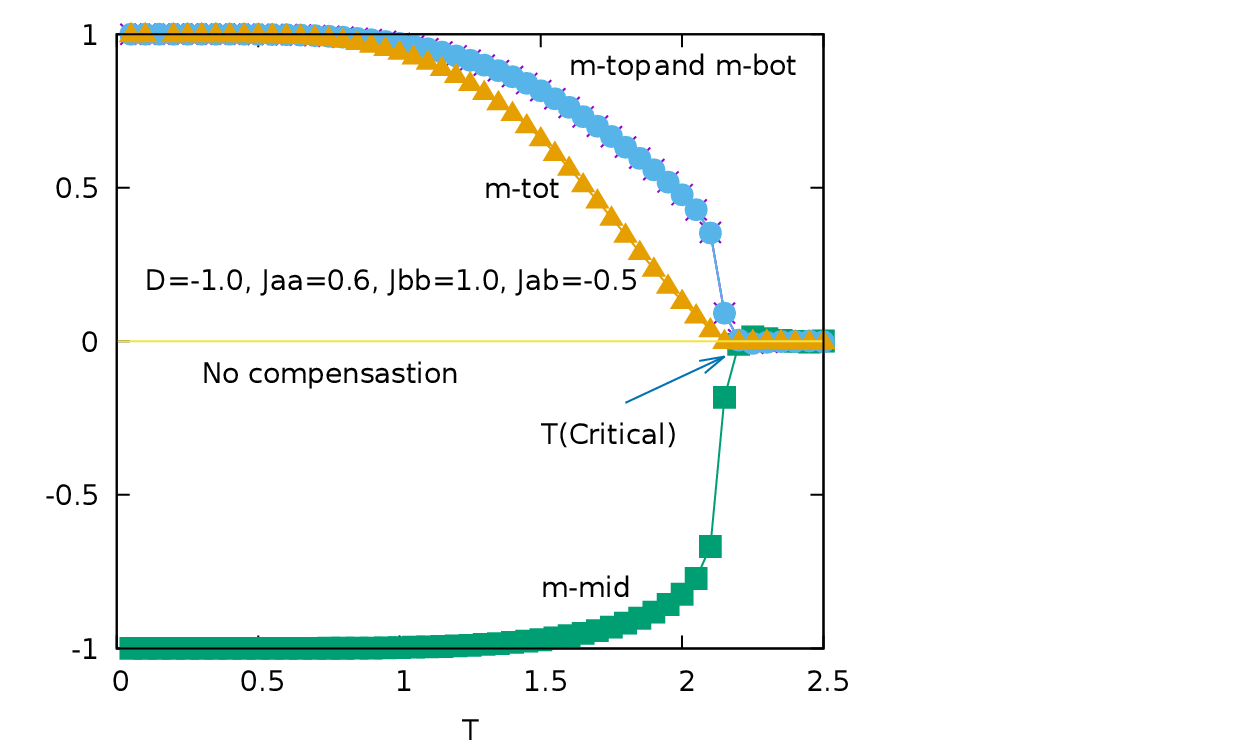}}
\\
	(c)
\resizebox{7cm}{!}{\includegraphics[angle=0]{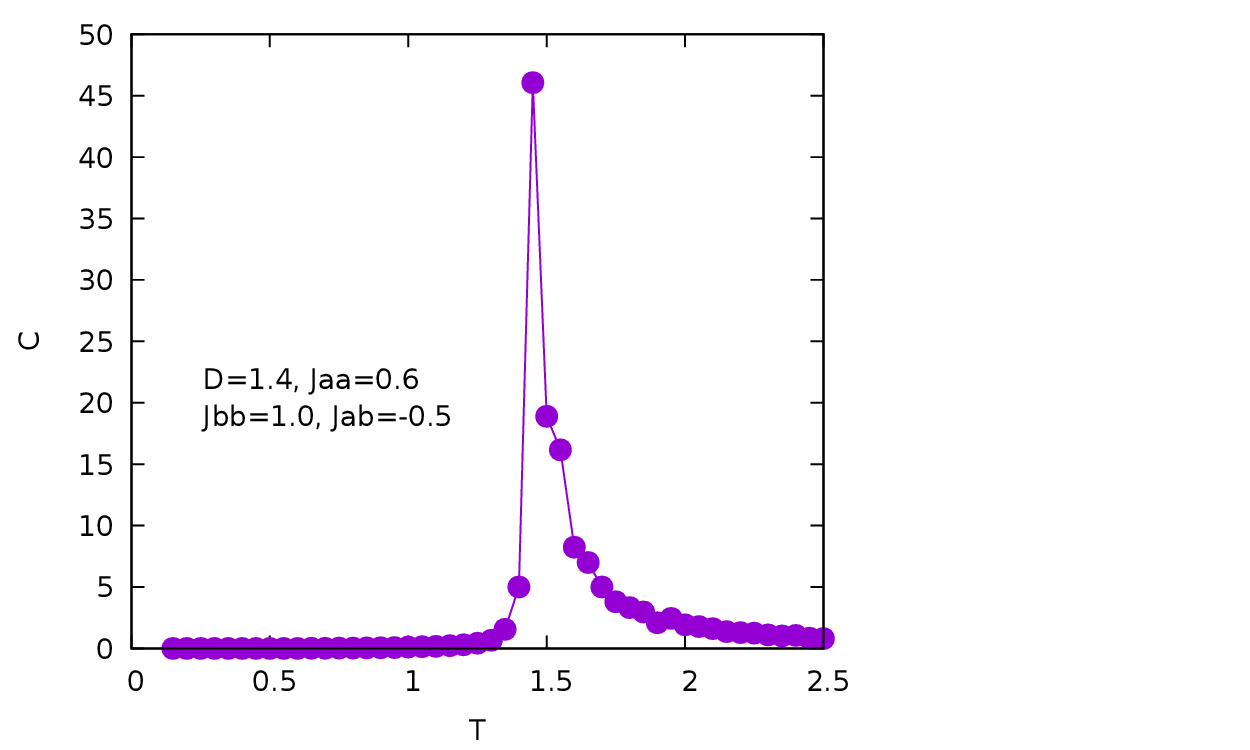}}
	(d)
\resizebox{7cm}{!}{\includegraphics[angle=0]{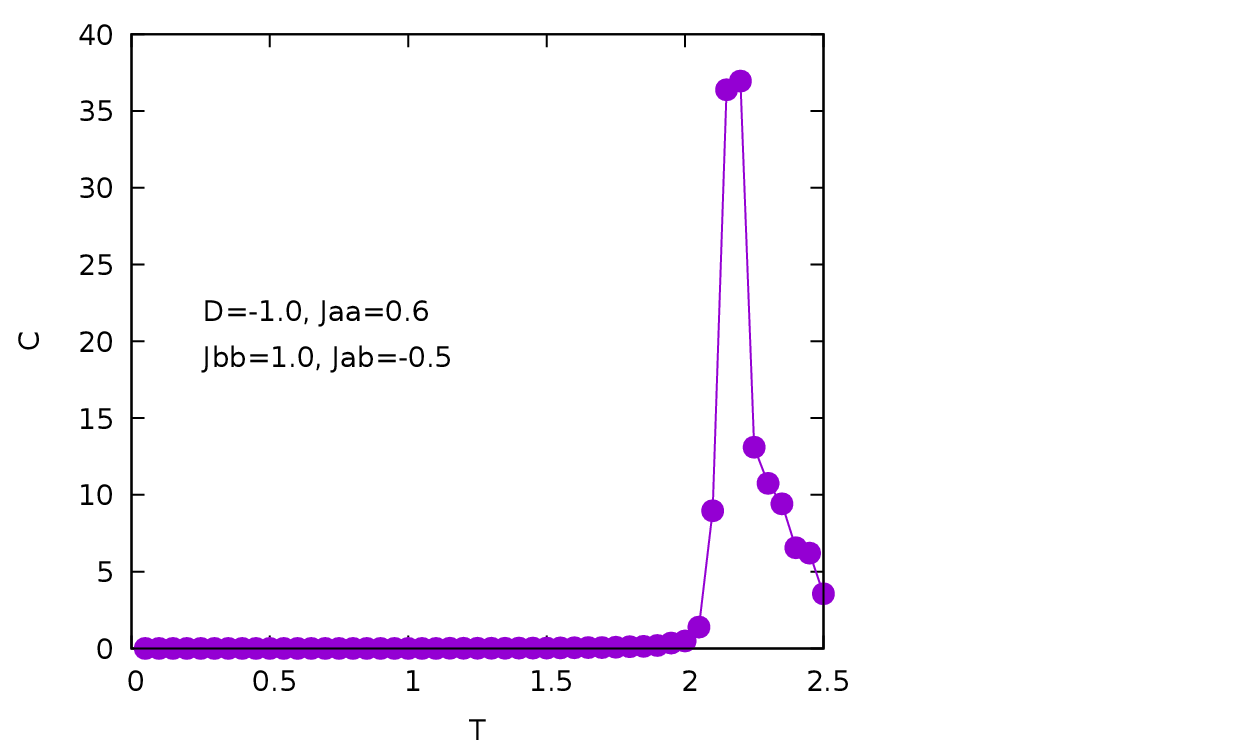}}
          \end{tabular}
\caption{The sublattice magnetisations of different layers and the total magnetisation are plotted against the temperature. The corresponding susceptibilities are
also plotted against the temperature of the system.
}

	\label{mT0.6}
\end{center}
\end{figure}

{\it What would happen if $J_{aa}$ is moderately higher, 
say $J_{aa}=0.6$?} In this case,
for negative anisotropy $D$ no compensation was observed. However, it
appears for positive anisotropy $D$.
In BC model the compensation is observed only for some combinations of values of
$D$ and $J_{aa}$. 
 Fig-\ref{mT0.6} shows such 
a comparison. For $D=1.4$, the compensation was observed 
(Fig-\ref{mT0.6}a).
The compensation was not found for $D=-1.0$(Fig-\ref{mT0.6}b).
In both cases, the critical temperatures $T_{critical}$ were estimated 
from the positions of maxima of the susceptibilities $C$ (Fig-\ref{mT0.6}c
and Fig-\ref{mT0.6}d). It may be noted that for relatively weaker $J_{aa}=0.2$,
the compensation appears for the entire range of values of the anisotropy
$D$. On the other hand, for relatively stronger $J_{aa}=0.8$, the compensation
was not observed at all. In this case, stronger ferromagnetic interaction
dominates over the role of the magnitude of $D$. Each sublattice provides the magnetisation
of almost equal magnitude. In between these two limits, for relatively
moderate value of $J_{aa}=0.6$, the compensation is observed for the
positive values of $D$. According to the form of BC Hamiltonian, positive
$D$ would increase the probability of having $S_i^z=0$ in the system which
effectively reduces the chance of having the configurations of all the spins 
(in a particular sublattice) to become parallel
(by ferromagnetic interaction $J_{aa}$). As a result, compensation is 
favoured. One has to keep in mind that compensation mechanism in BC model is 
controlled jointly by the anisotropy ($D$) (which provides additional degrees of freedom
of spin $S_i^z=0$) and $J_{aa}/J_{bb}$. The compensation
in BC model would be favoured for large $D$ and small $J_{aa}/J_{bb}$.

\begin{figure}[h]
\begin{center}
\begin{tabular}{c}
        \resizebox{7cm}{!}{\includegraphics[angle=0]{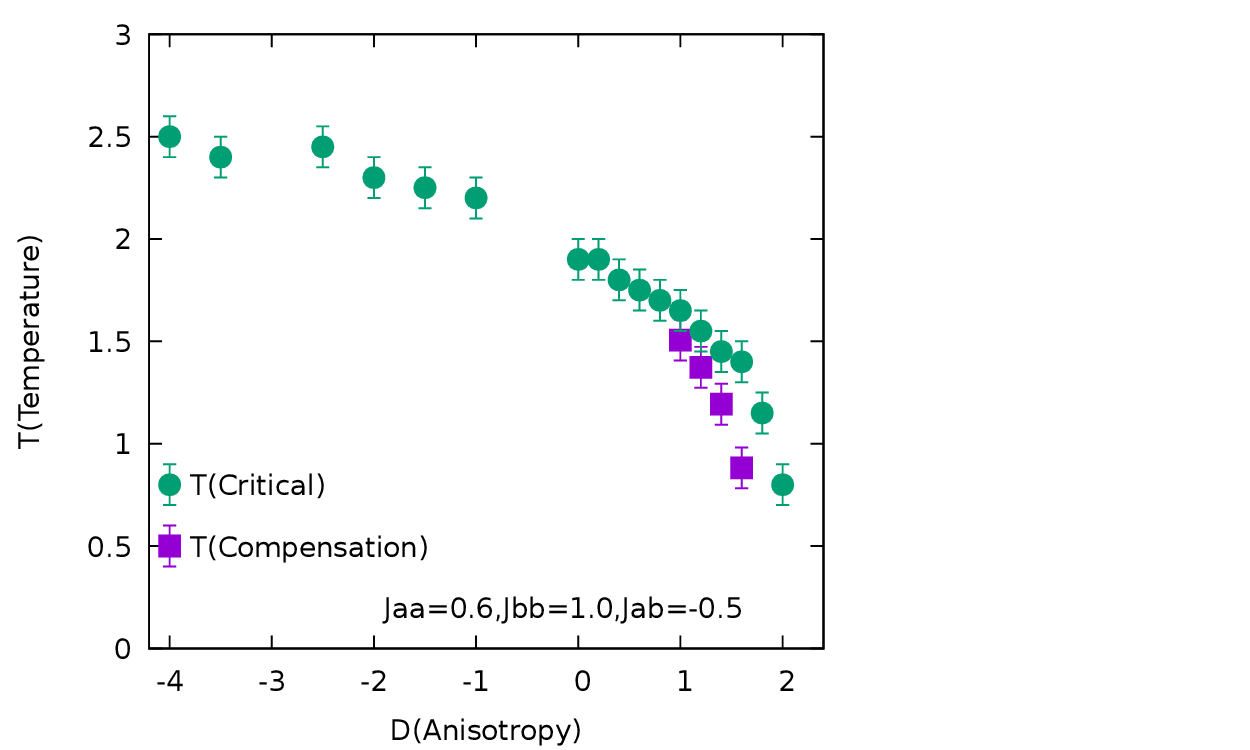}}
        
          \end{tabular}
	\caption{Phase diagram in the D-T plane. Here, $J_{aa}=0.6$,
	$J_{bb}=1.0$ and $J_{ab}=-0.5$.}
	\label{phs0.6}
\end{center}
\end{figure}

By estimating the critical temperature from the susceptibility $C$ and the
compensation temperature $T_{compensation}$ from the linear interpolation
near the change of sign of total magnetisation $m_{tot}$, the comprehensive
phase boundary was be obtained. Such a phase boundary was shown in
Fig-\ref{phs0.6}. It may be noted here that the compensation could be 
found only for positive values of $D$. A very narrow region 
bounded by the boundaries of $T_{compensation}$ and $T_{critical}$
was observed. The similar kind of phase boundary, having a meeting point
of the lines of critical and compensation temperatures, could be observed
also for slightly lower value of $J_{aa}$ where the meeting point may
be shifted towards negative value of $D$.

\begin{figure}[h]
\begin{center}
\begin{tabular}{c}
	(a)
\resizebox{7cm}{!}{\includegraphics[angle=0]{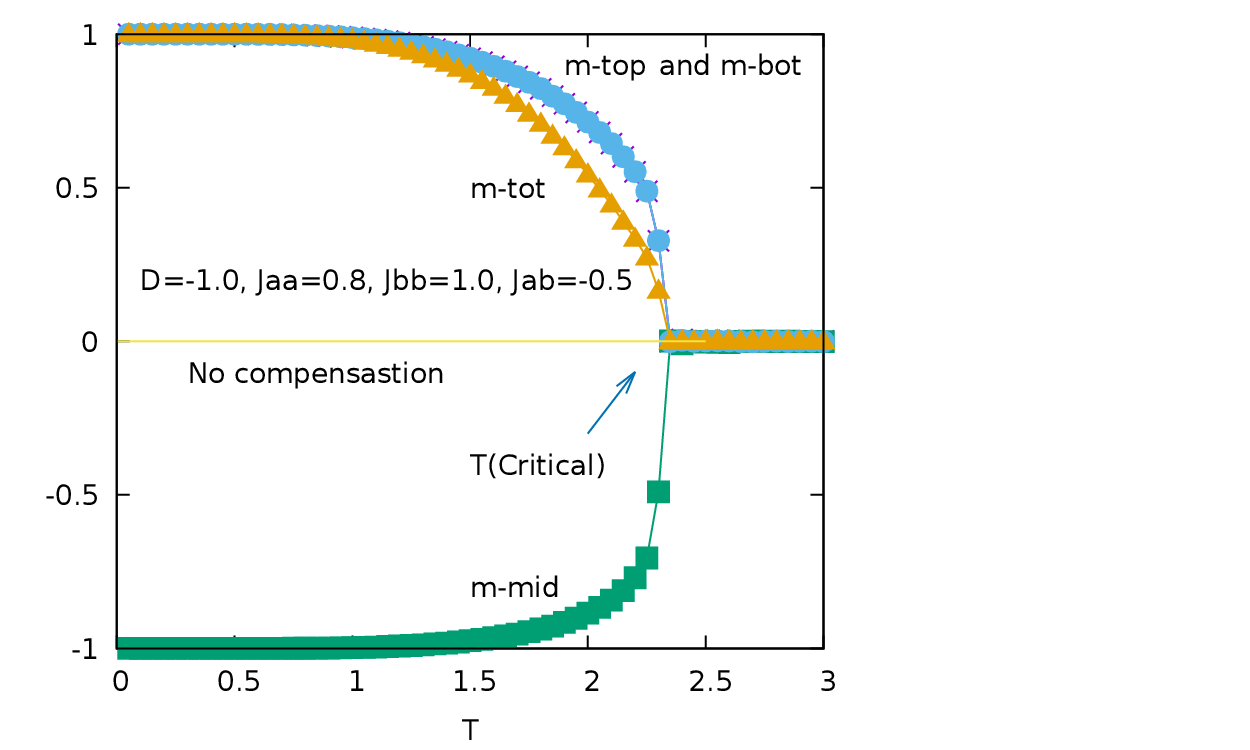}}
	(b)
\resizebox{7cm}{!}{\includegraphics[angle=0]{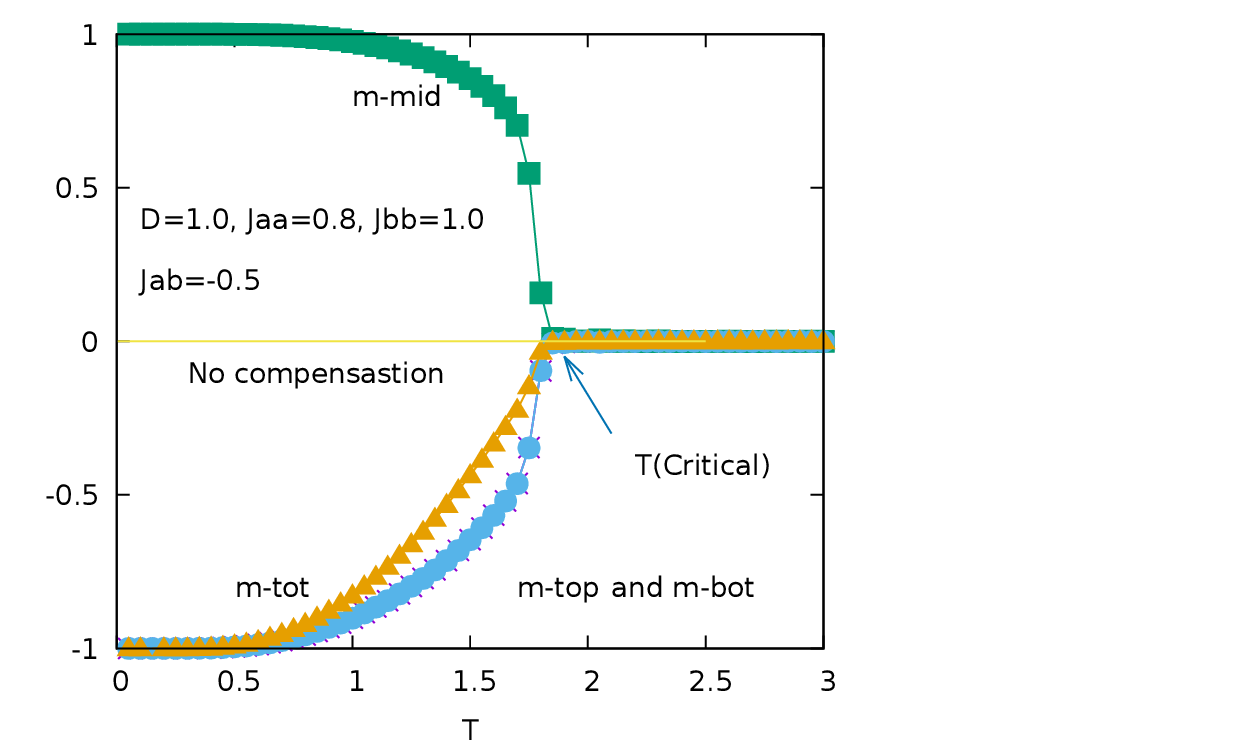}}
\\
	(c)
\resizebox{7cm}{!}{\includegraphics[angle=0]{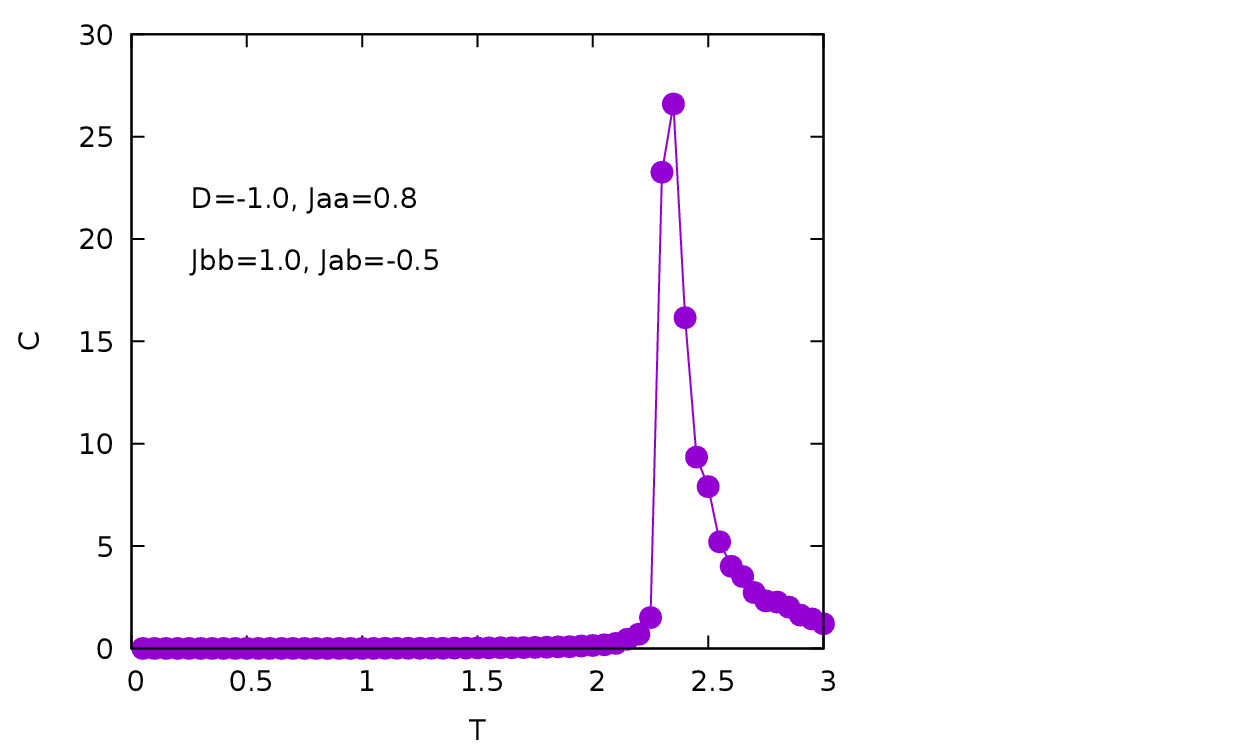}}
	(d)
\resizebox{7cm}{!}{\includegraphics[angle=0]{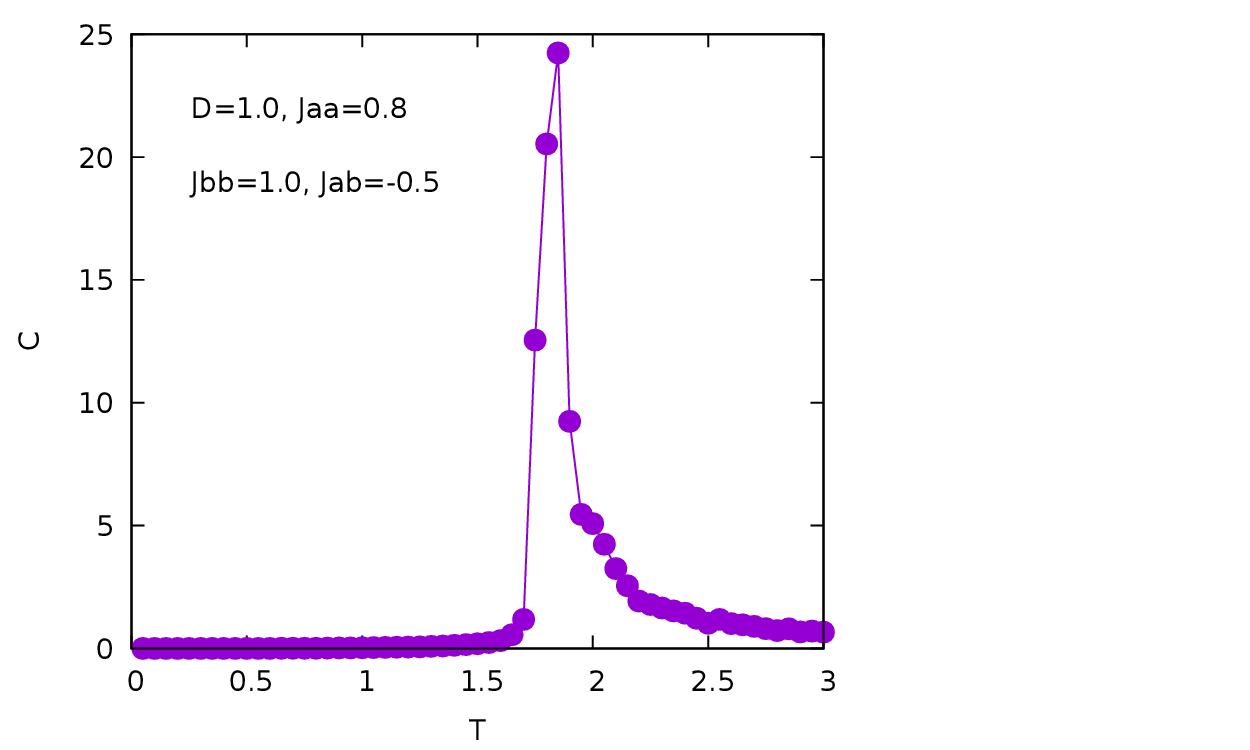}}
          \end{tabular}
\caption{The sublattice magnetisations of different layers and the total magnetisation are plotted against the temperature. The corresponding susceptibilities are
also plotted against the temperature of the system.
}

	\label{mT0.8}
\end{center}
\end{figure}

\begin{figure}[h]
\begin{center}
\begin{tabular}{c}
        \resizebox{7cm}{!}{\includegraphics[angle=0]{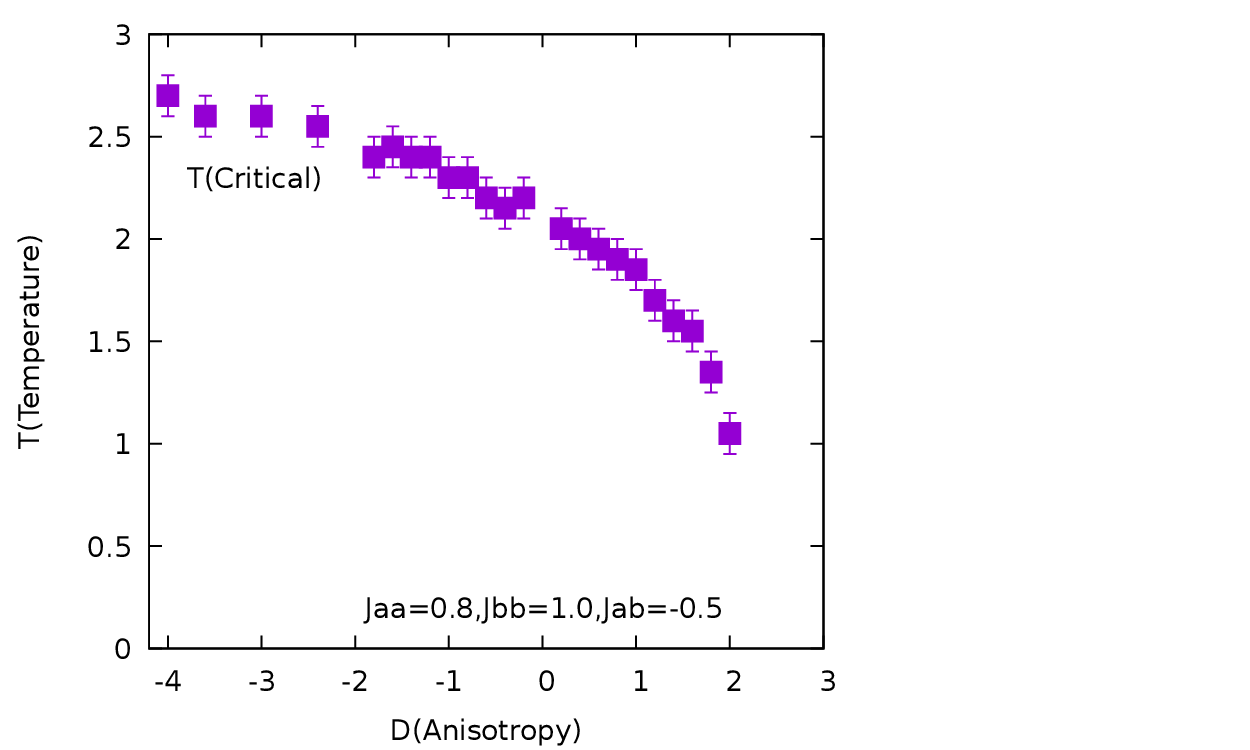}}
        
          \end{tabular}
	\caption{Phase diagram in the D-T plane. Here, $J_{aa}=0.8$,
	$J_{bb}=1.0$ and $J_{ab}=-0.5$. No compensation is observed here.}
	\label{phs0.8}
\end{center}
\end{figure}

{\it Can one expect to observe the compensation for very high value of 
$J_{aa}=0.8$ ?} Fig-\ref{mT0.8}a and Fig-\ref{mT0.8}b show the variations
of sublattice magnetisation and the total magnetisation as funtions
of the temperature. From these plots no compensation is observed. 
Due to the relatively strong ferromagnetic interaction $J_{aa}$, almost all
the spins in any sublattice becomes parallel which creates a situation of having
no compensation.
Only
the critical temperatures can be estimated from the temperature variations
of the susceptibility $C$ (shown in Fig-\ref{mT0.8}c 
and Fig-\ref{mT0.8}d). The comprehensive phase boundary 
(only for $T_{critical}$) was drawn and 
shown in Fig-\ref{phs0.8}. The set of values of the chosen interaction parameters is incapable of giving any compensation in this case.

\vskip 1cm

\noindent {\bf IV. Summary}

 The equlibrium properties of Blume-Capel trilayered magnet
have been studied by Monte Carlo simulation with Metropolis single spin
flip algorithm. The A-B-A type of trilayer is considered, where the
intralayer ferromagnetic interaction strength 
of the middle (B say) layer is $J_{bb}=1$. The other
two layers (bottom (A)  and top (A) say) have intralayer ferromagnetic 
interaction strength $J_{aa}$. The interlayer antiferromagnetic strength
is $J_{ab}$. For fixed values of $J_{bb}$=1 and $J_{ab}=-0.5$, the 
sublattice magnetisation was studied as function of temperature. 
The critical temperature was found from the maximum of the susceptibility
and the compensation temperature was determined from the linear 
interpolation of the two points where the total magnetisation changed sign.
The critical temperature and the compensation temperature were
studied as function of the anisotropy $D$ for different papermeter
values of the relative interaction strengths $J_{aa}/{J_{bb}}$, namely
weak, moderate and high.
For a range of values of the strength of anisotropy $D$, the weak
relative interaction $J_{aa}/{J_{bb}}=0.2$ 
shows compensation behaviours. 
In this range, the difference between the critical temperature
and the compensation temperature is quite high.
The
moderate $J_{aa}/{J_{bb}}=0.6$ value of relative interaction, shows 
compensation for positive values of the anisotropy only. 
In this case, the difference between the critical and compensation
temperatures are very small.
The compensation was not observed (in the wide range of 
values of $D$)
for high value of $J_{aa}/{J_{bb}}=0.8$. 

\begin{figure}[h]
\begin{center}
\begin{tabular}{c}
        \resizebox{7cm}{!}{\includegraphics[angle=0]{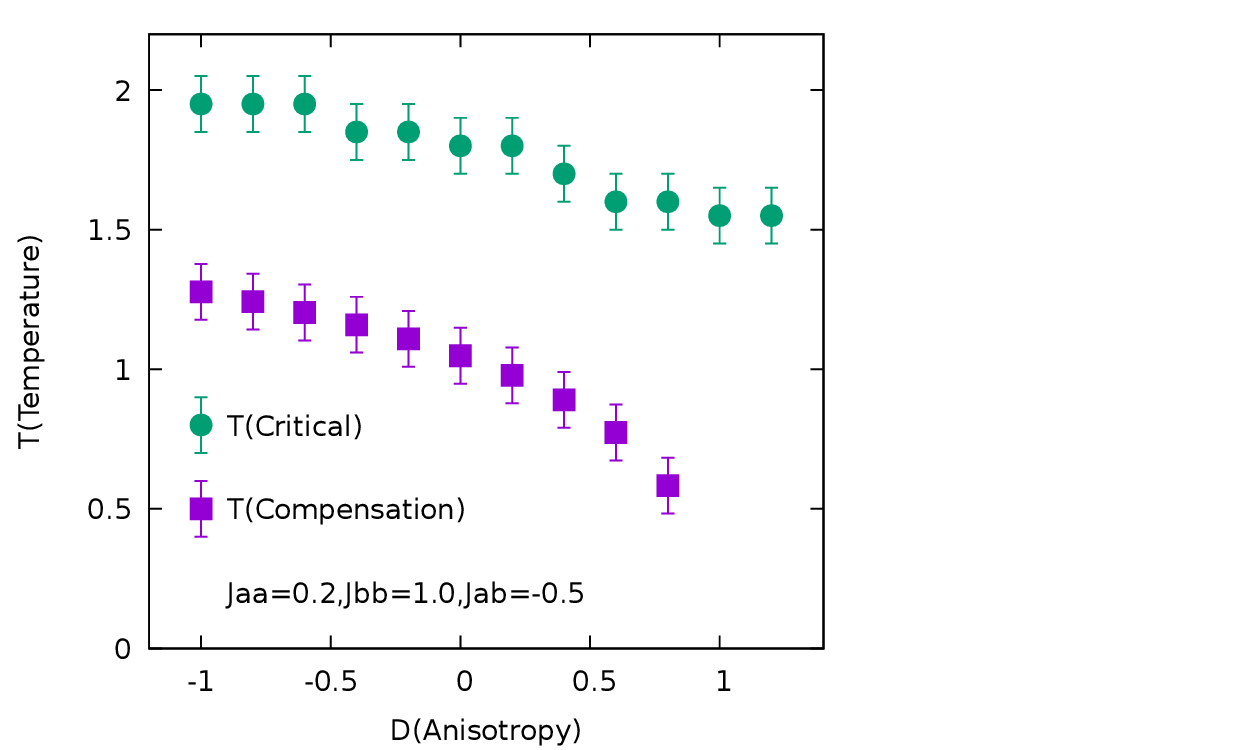}}
        
          \end{tabular}
	\caption{Phase diagram in the D-T plane. Here, $J_{aa}=0.2$,
	$J_{bb}=1.0$ and $J_{ab}=-0.5$.}
	\label{phs0.2}
\end{center}
\end{figure}

Why does the critical temperature decrease as the anisotropy $D$ increases (see
Fig-\ref{phs0.2}, Fig-\ref{phs0.6} and Fig-\ref{phs0.8}) ? In the Hamiltonian,
the term responsible for the single site anisotropy is $+D\sum_i (S_i^z)^2$ (note
the positive sign). For negative anisotropy, if the magnitude $|D|$ decreases
(in a sense $D$ increases in number line), the
spin flip (from $S_i^z=+1$ to $S_i^z=-1$) becomes more probable. This reduces the
critical temperature. On the other hand, if $|D|$ increases for positive $D$, the possibility of having
$S_i^z=0$ is more which leads to the reduction of the critical temperature. 

Why does the compensation disappear for large $J_{aa}$ ? In the present study the 
compensation is not observed for $J_{aa}=0.8$. The compensation is basically
the disappearnce of total magnetisation even with nonzero sublattice magnetisation.
It is a balancing condition which leads to the zero total magnetisation for a
special combination of the values of sublattice magnetisations.
At the critical temperature, the total magnetisation also vanishes. But each
sublattice magnetisation also vanishes there. If the relative strength of 
interactions $J_{aa}/J_{bb}$ in this A-B-A structure is close to unity, the
trilayered system forms a layered antiferromagnetic structure. Where each
layer is almost fully magnetised. But the direction
of spins are opposite in B layer than that in A layer due to interlayer antiferromagnetic interaction. This cannot lead to
compensation, since total magnetisation remains nonzero everywhere below the
critical temperature. However, if the relative strength is low enough, the
absolute value of the magnetisation of each layer are significantly different.
This situation has a possibility of having vanishing total magnetisation
(without nonzero sublattice magnetisation) at any finite temperature below
the critical temperature.  

This article is an effort to study the behaviours of compensation temperature and
critical temperature as functions of single site anisotropy in the ${\bf S=1}$ Blume-Capel trilayered (A-B-A type) model. The complicated phase diagarm of the system, depending on the relative
interaction strength and the single site anisotropy, may be useful to design
the magnetocaloric devices. This study is an appeal to the technologists to 
check in magnetic materials (by changing the anisotropy), how to maximize the
coercivity (close to compensation temperature). To study the compensation in
another kind of trilayered (A-A-B type) BC model would be interesting.

\vskip 1cm

\noindent {\bf V. Acknowledgements:} Author would like to acknowledge the
FRPDF grant provided by the Presidency University.

\newpage

\begin{figure}[h]
\begin{center}
\begin{tabular}{c}
\resizebox{9cm}{!}{\includegraphics[angle=0]{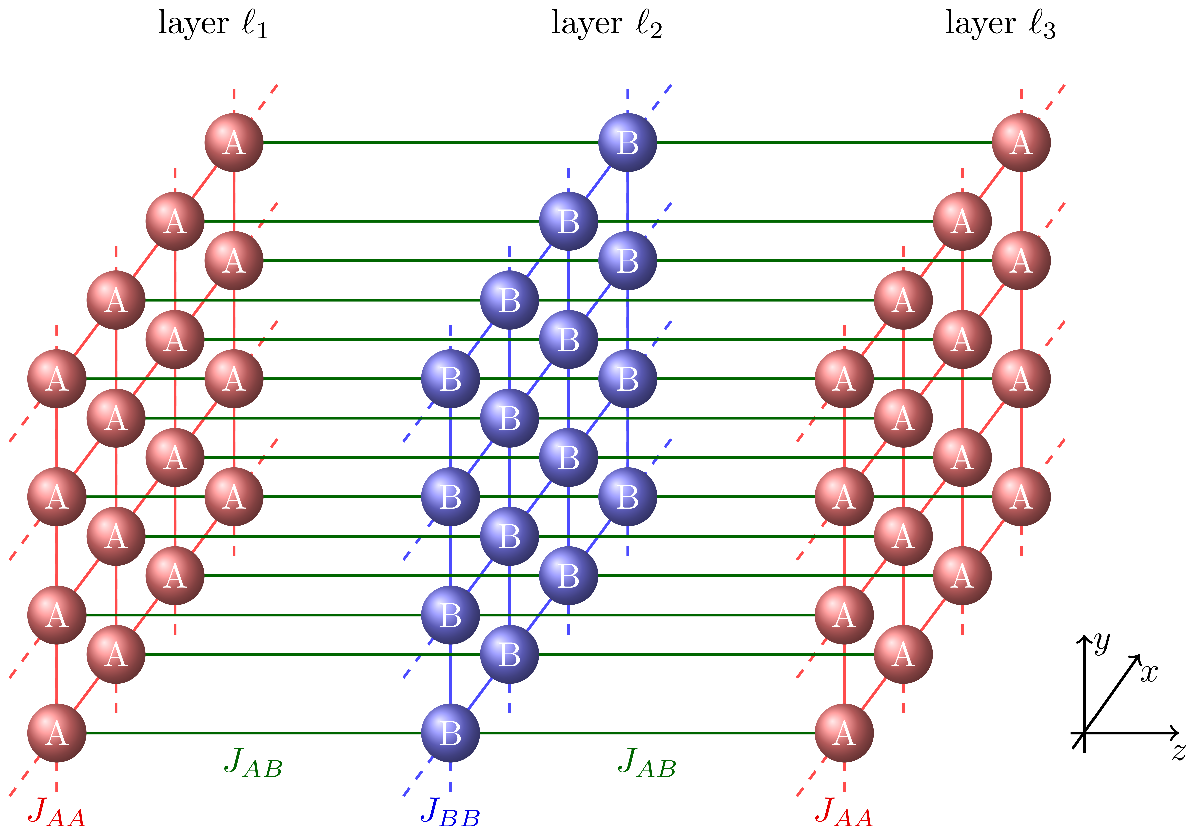}}
          \end{tabular}
\caption{
	The geometric structure of the system of trilayerd (A-B-A)
	magnetic model. Collected from I. J. L. Diaz 
	and N. S. Branco,
	cond-mat:1711.10367.
}

	\label{lattice-geometry}
\end{center}
\end{figure}
\newpage

\newpage

\newpage

\newpage

\newpage
\newpage

\newpage
\end{document}